# Children's Expectations, Engagement, and Evaluation of an LLM-enabled Spherical Visualization Platform in the Classroom

Emelie Fälton[1] , Isabelle Strömstedt[1] , Mathis Brossier[1] , Andreas Göransson[1] , Konrad Schönborn[1,4] , Amy Loutfi[1,2] ,
Erik Sunden[1] , Mujtaba Fadhil Jawad[1] , Yadgar Suleiman, Johanna Björklund[3] , Mario Romero[1] ,
Anders Ynnerman[1] , Lonni Besançon[1]

[1] Media and Information Technology, Linköping University, Sweden
[2] Örebro University, Sweden, [3] Umeå University, Sweden
[4] Department of Behavioral Sciences and Learning (IBL), Linköping University, Sweden

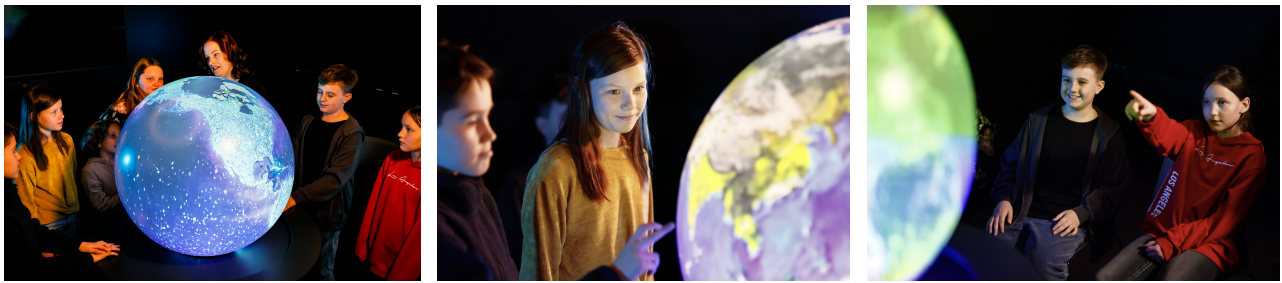

**Figure 1: Teaser.** Our LLM-enabled spherical visualization platform across different usage contexts: (left) a facilitated session in which a pedagogue demonstrates the system and explains earth-related datasets; (middle, right) child-led interactions in which children directly explore and query datasets.

**Abstract**
*We present our first stage results from deploying an LLM-augmented visualization software in a classroom setting to engage primary school children with earth-related datasets. Motivated by the growing interest in conversational AI as a means to support inquiry-based learning, we investigate children's expectations, engagement, and evaluation of a spoken LLM interface with a shared, immersive visualization system in a formal educational context. Our system integrates a speech-capable large language model with an interactive spherical display. It enables children to ask natural-language questions and receive coordinated verbal explanations and visual responses through the LLM-augmented visualization updating in real-time based on spoken queries. We report on a classroom study with Swedish children aged 9–10, combining structured observation and small-group discussions to capture expectations prior to interaction, interaction patterns during facilitated sessions, and children's reflections on their encounter afterward. Our results provide empirical insights into children's initial encounters with an LLM-enabled visualization platform within a classroom setting and their expectations, interactions, and evaluations of the system. These findings inform the technology's potential for educational use and highlight important directions for future research.*

**CCS Concepts**
• **Human-centered computing** → **Empirical studies in visualization;** *Interaction devices;*

## 1. Introduction

Interactive visualization tools have long played a central role in science centers, where hands-on engagement spark curiosity and support learning in STEM fields among young audiences. As an example of these technologies, spherical displays have gained growing interest for their ability to present planetary and astrophysical phenomena in intuitive and visually immersive ways [BEB*15, GKD10, SUAO*14]. Research shows that such displays improve conceptual understanding of spatial and temporal processing, particularly when combined with active facilitation by educators [HH15]. In this study, we explore how children engage with and interpret an LLM–mediated conversational interface that supports visioverbal (linked verbal and visual dialogue) around Earth and solar system visualizations, displayed on a spherical display (hereby referred to as the "LLM-enabled visualization platform").

The study is conducted within *TellUs – the Talking Planet*, a Swedish research project involving five science centers that explore how spherical displays visualizing planet Earth, augmented with LLM-based conversational interfaces, can engage children in STEM and environmental questions during outreach activities. The project brings the spherical displays into schools directly, where children interact with the system through a visioverbal mode in which spoken dialogue is closely linked to dynamic planetary visualizations, giving tangible form to the idea of a "talking planet."

Within *TellUs*, the spherical form is a conceptual and pedagogical element. The decision to use a spherical display is motivated by its capacity to render Earth data in alignment with the planet's own form, supporting interpretations grounded in planetary scale and materiality [BEB*15, SUAO*14]. Related studies further suggest that globe-based visualizations can afford embodied interaction, encourage movement around the display, and create a shared focal point for collaborative exploration and discussion [STJ*21]. This rationale is further strengthened by our previous work in this domain (see Figure 1), where a large spherical display was used to communicate earth-related datasets and climate change science to visitors of a science center [BSS*22], benefiting from the form factor of the display to maximize the engagement of large-audience groups [SSE*22, WSB15, VWB*14]. Interactions with spherical displays have also been studied [STJ*21] in other science centers.[†]

Our visioverbal AI implementation can navigate different datasets and provide linked spoken and visual explanations. We are also experimenting with anthropomorphization of earth by letting the AI assume the identity of our planet. This LLM interaction, where dynamic visualizations are coupled with conversational dialogue, represents an emerging approach in learning contexts [JIB*25]. Due to the outreach context of *TellUs*, we develop our LLM-enabled visualization platform to be used in the context of a classroom. We aim to evaluate how an LLM-augmented data visualization software can perform in such a learning context with primaryschool children. With our data collection and analysis, we aim to better understand children's expectations, engagement, and evaluation of LLM-enabled visualization platforms.

Previous research has focused either on LLM-based conversational guides in informal learning settings [CPW*19], or on children's perceptions of AI technologies such as smart speakers [AR23]. We extend this body of work by examining how children experience and make sense of our LLM-enabled visualization platform when it is introduced in a formal classroom context. We argue that our findings will help us better derive specific design factors to consider for future work in this context. As the intervention represents the first classroom deployment of the LLM-enabled visualization platform, the primary aim was not theory generation or hypothesis testing, but to capture children's expectations, interaction practices, and reflections in situ. Drawing on user studies with school children ages 9–10, we present insights from the first classroom session in which children were invited to interact with the platform, thereby allowing us to examine how it functions when integrated into everyday school practices. By foregrounding children's encounters with the platform as an interactive visualization tool, the study contributes to an empirical foundation for examining the interplay between technical affordances and pedagogical design [FBP04]. To summarize, we contribute to the following research questions:

---

[†] see e.g., https://www.miraikan.jst.go.jp/en/

- What expectations do children have regarding interacting with an LLM-enabled visualization platform before using it?
- How do children engage and interact through spoken dialogue with the platform during facilitated classroom sessions?
- How do children experience and evaluate their interaction with the platform after the classroom sessions?

This study contributes to educational visualization research [SB24] in several ways. First, it contributes further results on children's initial encounters with an LLM-enabled visualization platform when introduced in a classroom setting, extending prior work that has primarily examined spherical displays in informal learning environments [OSH19, SUAO*14]. Second, by analyzing children's expectations prior to interaction, their engagement during facilitated sessions, and their reflections afterward, the study offers a detailed account of how children experience and make sense of LLM interaction in a classroom context. Third, the study identifies design-relevant affordances and constraints of integrating LLMs with shared visualization systems, including challenges related to conversational turn-taking, explanation, complexity, and alignment between spoken responses and visual feedback. Overall, the paper contributes conceptually by positioning LLM-enabled visual interaction as an emerging pedagogical configuration.

## 2. Related work

We present research exploring the educational value of interactive visualizations and the use of LLM interfaces aimed at augmenting user experiences in data analysis and exploration.

### 2.1. Spherical Displays

Research on educational visualizations in formal school settings for young children is extensive and spans a wide range of interactive technologies, including student-generated visual representations, visual literacy, and the use of digital tools for learning, e.g., [ARC*17, PSN24, PWD*24]. Within this extensive work, studies of approaches such as *Slowmation* illustrate how interactive visualization can stimulate creativity, foster a sense of ownership, and support conceptual development among young learners [BMH13]. These studies are a small slice of a larger field that, so far, gives limited insight into how immersive visualization, especially spherical displays, work in everyday classrooms. We review prior findings.

Prior studies have examined spherical displays as interfaces for exploring Earth science and planetary phenomena through interactive visual representations. Unlike flat screens, physical globes are not subject to distortions when presenting spherical objects. Hence, they offer a perceptually coherent representation of Earth [BEB*15, SUAO*14]. Most studies have been conducted in informal learning environments, particularly science centers, where spherical displays are embedded within guided narratives [GKD10, HH15]. While this body of work demonstrates the potential of immersive visualization to foster curiosity, engagement, and conceptual understandings of Earth, it also reveals notable limitations. The form factor lacks a single "front," which intrinsically creates both shared and private interaction zones based on each user's viewing angle and position [STJ*21]. Most studies focus on interactions with spherical displays in exhibition settings, leaving open questions about how such devices function when integrated into everyday educational practices, such as classrooms [OSH19, SUAO*14].

As users self-organize in space, they naturally coalesce into *F-formations* (Kendon's term for spatial configurations that sustain a focused interaction space [Ken10]). F-formations have a direct influence on collaborative learning dynamics: a field study in a science museum found that the physical spacing of group members around a multi-touch sphere was strongly linked to how they collaborated, with tightly clustered groups engaging in coordinated discussion and widely spaced groups tending toward independent exploration [STJ*21]. Taken together, this body of research points to the potential of immersive globe-based displays, but provide limited insight into how visualization systems, such as spherical displays, translate into classroom environments where interaction is collective, spatially constrained, and shaped by existing pedagogical routines. In addition to this contextual gap, previous educational visualization research has largely examined visual interaction in isolation from conversational or dialogic systems.

Although recent advances in LLMs suggest that such systems can interpret and explain visualizations [MKB*25], empirical studies investigating their integration into educational visualization interfaces remain scarce [BS25]. Addressing these gaps, we investigate how an AI-enabled spherical display is introduced and experienced in a classroom setting. Rather than evaluating learning outcomes in isolation, we focus on children's expectations prior to interaction, their engagement during facilitated sessions, and their reflections afterward. By examining these dimensions, we contribute to educational visualization research by offering empirical insights into how LLM interaction, where dynamic visualizations are tightly coupled with conversational dialogue, functions when immersive display technologies are embedded in everyday school practices.

**2.2. LLM-augmented Visualizations**

The field of visualization is currently exploring how LLMs can expand the explanatory capabilities of interactive visualizations. In museum contexts, conversational LLMs have the potential to enhance interactive exhibits by improving visitor's engagement and immersion. This idea has been developed through the shared and co-created imagined futures of 47 museum guides [CPW*19]. Moreover, it has been deployed and tested in museum environments: for instance, Ghosh et al. [GRS21] implemented a low-cost virtual museum guide that combines a domain-specific chatbot with IoT infrastructure to enable partial control of some of the exhibits, mimicking what a pedagogue would provide in visit contexts. In their test, the authors found that visitors first engage quite strongly with the agent, but that their satisfaction decreases with time, potentially due to current technical limitations of LLMs. These findings are further supported by the literature review conducted by Li et al. [LZWO24], which highlights the potential of LLM-powered agents to provide personalized educational tools, as well as the technical challenges that currently impair interaction (e.g., response time and query interpretation). The potential of LLM agents to provide a tailored experience in conversation interaction is further highlighted in the literature survey by Diederich et al. [DBMK22]. While we share the value of trying to enhance each visit and tour to a museum through an LLM-powered agent, our work instead focuses on providing a single LLM agent for a single exhibit containing interactive visualizations, in particular for school-aged children.

In the domain of children's perceptions of interacting with LLMs the work of Andries and Robertson [AR23] is particularly valuable. They investigated the perception of smart speakers with 166 children, ages 6 to 11 years. They found that children tended to overestimate the cognitive abilities of smart speakers but that this kind of interaction made them more curious about AI and its capabilities in general. More focused on visualization, LLMs are often used to empower visually-impaired users. Manzoni et al. [MMA*25] propose using LLM conversations to increase accessibility of interactive maps. Work by Gorniak et al. [GKWK24] has investigated the possibility to enable users to rely on natural language for querying visual data trends and to provide more contextual information about interactive visualizations while providing more accessibility to visually-impaired users. Similarly, Reinders et al. [RBZ*25] investigate conversational agents to increase accessibility for data analysis. While relevant, these works do not directly focus on educational interventions in a classroom context with children.

The nearest related work is by Jia et al. [JIB*25]. The authors coupled an LLM to a visualization allowing for natural-language data queries triggering navigation, filtering, and exploration while providing contextual knowledge. The authors studied how researchers in museum pedagogy value this system. They highlighted that the LLM agent can be used by visitors on their own by answering their questions and update the visualization in real time. It can also be used together with a pedagogue. In this case, it helps reduce the pedagogue's interactive workload on the exhibits by listening and updating the visualization in real time. It can also provide answers to contextual questions that the pedagogue may not have. Our work is directly inspired by this: we aim to test in a classroom setting how this combination of an LLM and a visualization can be used by children for educational purposes.

**3. Methodology**

This study employs an exploratory qualitative approach combining structured observation and semi-structured interviews, alongside an iterative, empirically grounded analysis. In this section, we describe the apparatus, the context of the classroom interventions, and the collection of data as well as the analytic approach.

**3.1. Apparatus**

The spherical display (Figure 1) used as part of our LLM-enabled visualization platform is a PufferSphere[‡], specifically a 760 mm diameter touch-responsive globe. It is mounted on a company-provided metal housing that contains the projector, a computer, speakerphones, and a power supply. The housing contains a bright, 4K projector with a fisheye lens projecting upwards into the hollow, translucent sphere coated with a diffusion surface. The projector is sufficiently bright for indoor viewing under ambient light, and the device software supports display of 360-degree images and videos, as well as smaller 2D images and infographics. The visualization is controlled by a speech-to-speech LLM (gpt-realtime-2025-08-28 by OpenAI[§], which can interact with the visualization by changing the displayed dataset or rotating the globe visualization (Figure 2). In this setup, the LLM generates verbal explanations and visual linked context generated with the OpenSpace software [BAC*20, BBS*24]. The LLM speech synthesis voice used is gender-neutral and is described as "Neutral and Balanced". Touch interaction supports rotation via one-finger swipes or two-finger twists. The visualizations depict bodies in the Solar System, including planets, the Sun, and major moons. On Earth, several data layers are also available, including cloud cover, forest cover, surface temperatures, and country borders. Each dataset is available to the LLM, along with a brief description of its contents.

---

[‡] https://pufferfishdisplays.com/products/puffertouch/

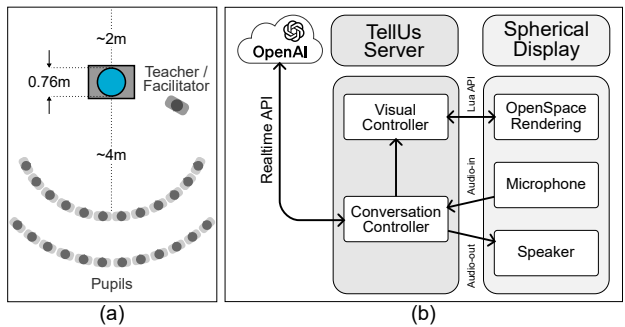

**Figure 2:** *System diagrams: (a) Bird's-eye view of the physical layout of the sphere in the classroom; (b) System architecture.*

### 3.2. The Classroom Interventions

To evaluate children's interactions with the LLM-enabled visualization, we conducted a study with two classes at a primary school in Sweden with children aged 9–10. The school was selected to reflect the *TellUs* project's focus on children aged 6–12 living in socioeconomically disadvantaged areas, a group that is often underrepresented in science center–based educational activities. This underrepresentation has partly been attributed to structural barriers, including limited access to transportation [GST23]. The present study focused on children aged 9–10 (grades 3–4). Within *TellUs*, content is adapted to different grade levels, with grades 3–4 addressing the use of AI. These interventions marked the first occasion the platform was tested with its target audience in its intended context of use. These test results are intended to inform further development of the platform, guide content creation, and support the design of further outreach activities. Testing in a classroom context was particularly important as it afforded observations of children interacting with the platform in a realistic learning environment, with its social dynamics and group behaviors.

The platform was positioned in a classroom with an open radius of roughly 2 meters, allowing participants to move freely around it (Figure 2). Pupils were seated on chairs arranged in a semicircle on one side of the platform. This configuration ensured all children could view the content clearly, while allowing an educator to move around the display without obstructing the view. The science center educator guided the sessions, curating interactions and supporting engagement with the display.

---

[§] https://platform.openai.com/docs/models/gpt-realtime

The study was based on two class sessions, each lasting approximately 90 minutes. Sessions followed a structured format to allow for comparability while enabling pupils to engage with both the LLM-enabled visualization platform and participate in reflective discussions. Each session was structured in three parts: an introductory segment conducted with the whole class, followed by two activity segments in which the class was divided into two smaller groups. During the introductory session, a researcher outlined the purpose and structure of the session and its two parts for 5 minutes. This was followed by a 20-minute presentation by the science center educator who facilitated the activity segment. She introduced the thematic content designed for pupils in grade 1-2, the Earth, to provide the children with an initial description of the platform and how it could be used, thus framing the activity conceptually and contextually prior to the activity blocks.

During the first activity block, each class was split into two groups with between 8-10 pupils in each group. Group 1 interacted with the LLM-enabled visualization platform, asking it questions about space and Earth and receiving visioverbal responses. A microphone captured questions directed to the LLM and was operated discreetly by an engineer to avoid interfering with the children's experience. Simultaneously, Group 2 participated in a small-group discussion led by a researcher and supported by an engineering student, focusing on pupils' expectations and perceptions of LLM-enabled visualization platforms. After 15 minutes, the groups switched activities, allowing all pupils to experience both interaction with the platform and engagement in reflective discussion. In the second activity block, Group 2 started with the small-group discussions. The group discussed their experiences interacting with the platform. Then, the groups switched activities again, ensuring that each child had the opportunity to both explore the applications and participate in discussions about their experiences. This structured rotation enabled the researchers to collect rich qualitative observations of pupil interactions and experiences.

### 3.3. Participants

A total of 33 children participated (19 girls and 14 boys). The participating classes were characterized by substantial linguistic and cultural diversity. Many pupils came from families with immigrant backgrounds, including origins in Arabic-speaking countries, Eastern Europe, and several African regions. Most children reported speaking two or three languages in their everyday lives. Permission to conduct the study was obtained from the participating school. Written informed consent was obtained from the legal guardians of all participating children, and the research was reviewed and approved by the Swedish Ethical Review Authority (Ethics Approval No. 2024-04416-01).

## 3.4. Collection of Data and Analysis

Across the two class sessions, researchers collected data onsite. Two researchers remained in the classroom with the LLM-enabled platform, writing detailed observations and hand-drawn sketches of the children's interactions with the platform, capturing spatial arrangements, interaction patterns and engagement. Simultaneously, two other researchers conducted semi-structured interviews with the other group of children in an adjacent room.

The interviews explored pupils' expectations prior to interacting with the system, including what they hoped it would do and how they imagined engaging with it, as well as their reflections afterward. These reflections covered what they liked, disliked, and suggest improving. Follow-up questions were posed as needed to clarify or expand responses. The interviewer asked questions based on a semi-structured interview guide (e.g., "What kinds of questions would you ask?", "What did you want it to talk about?", "What was it like to talk to the globe?", and "What did you think about the answers you received?"), with follow-up questions adapted to the ongoing dialogue with the children. All notes were taken in situ at a descriptive level. The resulting material corpus comprises structured observational notes documenting children's interaction with the platform, sketches capturing collective interaction and spatial dynamics around the display, and written notes from semi-structured interviews conducted pre and post interaction. The observations focused on collective interaction, turn-taking, verbal and non-verbal responses, and educator–child dynamics.

The analysis of the observation notes and sketches followed an iterative and collaborative process in which the research team collectively reviewed the collected material. The material was initially organized in relation to the study's three research questions, providing an empirically grounded analytical structure. Building on this framework, emerging themes and patterns were jointly discussed and guided further analytical attention beyond the initial research questions. This approach allowed the analysis to remain closely aligned with the exploratory nature of the study, while also supporting comparison across classes. The analysis foregrounds concrete interaction practices and articulated experiences, which may inform both further system development and future studies.

Children's expectations of interacting with the LLM-enabled visualization platform, were extracted by analyzing the interview notes from group discussions prior to pupils' interaction with the LLM. The analysis identified recurring themes in how the children articulated anticipated system capabilities, desired forms of interaction, and preferred content. To investigate children's interaction with the LLM-enabled visualization platform, the observation notes from the classroom sessions were analyzed in which each pupil was given the opportunity to ask questions and receive visual and verbal responses from the system. This analysis focused on how the interaction was structured and on the types of questions the children posed. Children's experiences in using the LLM-enabled visualization platform were analyzed based on the interview notes from post-interaction group discussions conducted immediately after the sessions. The analysis focused on how children evaluated the system in terms of usability, credibility, voice, and visual support.

## 4. Results

In the following section, identified themes are described and illustrated through selected examples from the obtained data.

### 4.1. Children's Expectations of Interacting with an LLM-enabled Visualization Platform

First, the children expressed both excitement and disbelief at the possibility of interacting verbally with the platform, responding with remarks such as "What? Is it true?" and "How?" When informed that they would be able to ask questions about space, many immediately articulated expectations related to space travels, seeing Saturn's rings, observing astronauts, and learning about the largest objects in the universe, the speed of light, and the extent of human exploration. The children wondered if they could pose questions beyond Earth and space, and several expressed disappointment when told by one of the researchers it was not possible. Cultural discourses surrounding artificial intelligence were also evident in the children's talk. For example, one child jokingly remarked that "AI is going to take over the world", reflecting familiarity with popular narratives and media representations of emerging technologies.

A second salient theme concerned children's desire to connect the platform's content to their everyday lives. 17 of the 33 pupils across the two sessions stated that they wanted to zoom in on the globe, view their own neighborhoods, explore cities, and even see themselves through services such as Google Earth or Google Street View. This was further evident when one child proposed a quiz about world cuisine, in which the platform would display an image of a dish and the audience would guess its country of origin. This suggestion prompted a lively discussion among the children about their own favorite foods, illustrating how they envisioned the platform as a means of linking global information to lived experiences. Children also expressed a desire to speak to the platform in multiple languages and suggested that either the AI should guess the language being spoken or that the children should guess the language spoken by the platform.

Beyond language, children expressed strong interest in both Sweden and the world. One child wished to see well-known Swedish landmarks, such as Skansen and the Vasa ship, both as a wreck under the sea and as a museum exhibit, while another wanted to explore rain forests, capitals, and famous places through images and videos. A recurring request was for more detailed Earth visuals, including oceans with fish and boats, shipwrecks, and people playing football. Some children proposed temporally oriented content, such as one child's request for a Big Bang visualization, another's interest in a planetary history timeline, and a third's wish to see representations of the future. Animals constituted another recurring interest, where four pupils discussed searching for different species, seeing where they are commonly found, viewing images, and reading short factual descriptions.

Although we introduced the activity as an opportunity to explore space, the children asked a substantial number of questions related to Earth and their immediate surroundings. In the classroom setting, these inquiries were predominantly educational in nature, in contrast to interactions observed when the same platform was demon-strated in a science center. Rather than engaging in playful or humorous requests, such as the "Mars taco recipe" requested by one child in the science center, the pupils' questions reflected expectations associated with formal learning situations. This finding indicates that both the classroom context and the facilitated structure of the intervention influenced how children framed their interaction with the LLM-enabled visualization platform.

### 4.2. Children's Interaction with an LLM-enabled Visualization Platform

Each session began with a brief introduction by the educator, followed by three demonstration questions, which were designed to illustrate the system's capacity to visualize graphical locations, respond to factual questions, and transition between earth and outer space. Following this, each child was invited to pose at least one question each to the platform. In the first session, the microphone was initially passed to pupils who raised their hands, but since some children dominated the interaction while others remained silent, subsequent sessions adopted a fixed order in which the microphone was passed along the rows. This structure ensured that all children participated and allowed pupils to plan their questions in advance, often resulting in brief follow-up questions.

The children's questions followed a loose but recurring trajectory. Interactions typically began with exploration of different planets in the solar system, continued with questions about space and the universe, and then returned to Earth, where children focused on countries, cities, and general world facts. While the specific questions varied, most concerned astronomy, geography, population and culture. The linguistic form of the questions also exhibited recurring patterns. Many questions were framed using superlatives and comparisons, such as "what is the biggest", "the furthest", and "the coolest" object in the universe (21% questions). Another common pattern involved inquiries about scale and distance, such as "how many people live in the USA", or "how far is the Moon from Earth" (12% questions). These questions frequently prompted LLM responses involving extremely large numbers difficult to conceptualize, and often expressed in terms such as "billions of billions". The LLM attempted to contextualize such information by providing comparisons or shifting to smaller-scale descriptions. In several instances, however, the generated responses exceeded children's observed level of comprehension, e.g., by expressing distances in light-years, referencing chemical elements, or providing information that was overly dense or only weakly connected to the original question. The rest of the questions were direct navigation queries (46%) and questions about facts or anecdotes (21%).

During the interaction, a discrepancy became apparent between the LLM's perceived dialogic capability, reinforced by its naturalistic voice, and the technical limitations of the prototype as a whole. While the LLM generally demonstrated strong performance in natural language understanding and child-directed interaction, its sensory input and interactive affordances differed fundamentally from those of a human interlocutor. First, speech recognition functioned reliably only when children spoke directly into the microphone. When a child varied the distance to the microphone mid-utterance, the system often captured only fragments of the question, leading the LLM to infer missing information, and in some cases, produce inaccurate interpretations. Second, the LLM struggled with conversational turn-taking. It responded immediately when the child stopped speaking, but could not distinguish between pauses for thinking from those signaling completion of a conversational turn. Likewise, the system lacked sensory cues for when to stop speaking. As a result, children who wished to ask follow-up questions were required to wait until the LLM had completed its response.

### 4.3. Children's Experiences of Interacting with the LLM-enabled Visualization Platform

The children articulated a wide range of reactions that reveal how they evaluated the conversational AI in terms of usability, credibility, and perceived confidence. Their comments reflected excitement and curiosity, alongside clear expectations regarding the system's limitations and how an AI should sound, respond, and behave. As in their expectations prior to the interaction, the children expressed a desire for stronger visual support, including more images, closer views, improved zoom functionality, clearer visual representations, and more dynamic visuals directly linked to their questions.

Twenty of the thirty-one children described the system as "easy to use", "fun" or "exciting". Several highlighted the novelty of being able to ask questions verbally and receive responses, with one pupil commenting that they had "never talked to an AI before". One child described it as resembling a conversation with "a real person", while others emphasized that the platform "showed interesting things", such as different planets. Some pupils explicitly referred to learning outcomes, where one child stated having "learned new things" and another that the platform "explained well". At the same time, ten pupils explicitly pointed to perceived limitations, such as the system occasionally misunderstanding questions, providing too complex or confusing answers, or contained unfamiliar vocabulary. One pupil noted a mismatch between the verbal and visual output of the system. For example, when asking about Mars, the system did not automatically display the planet unless explicitly prompted, which led to frustration and requests for more automated visual responses. Another child commented on delays in verbal responses or image loading, and five pupils expressed disappointment when the system responded with "I can't do that" to requests for unsupported content (e.g., black holes). One pupil summarized this limitation by remarking that the AI felt "born yesterday", indicating frustration with its perceived lack of understanding.

The system's voice emerged as a salient aspect of the experience. Eleven children described it as sounding "like an AI", often comparing it to familiar commercial assistants such as Siri or Google, while others characterised it as a hybrid between a human and a robot. One child noted the absence of breathing sounds or emotional variation, while another commented that "it just sounds happy". Additional observations included irregular pauses, shifts in tone, and moments in which the voice appeared to "glitch". In contrast, some children perceived the voice as more human-like, where one child stated that it "sounded like a real person", while another remarked that "it sounds like us", referring to its youthful tone. With respect to perceived gender, given that the LLM was implemented with a non-binary voice, one child described the voice as shifting between male and female. When asked about voice preferences, gendered patterns emerged with boys tending to prefer a male-sounding voice, whereas girls more often preferred a female-sounding one. Preferences regarding voice design varied considerably. One child stated that they did not want either a child-like voice or an overly robotic one. When asked about preferred alternatives, responses ranged from imaginative suggestions, such as a girl's voice described as a "sassy stepsister", or a teenager-like voice using slang and attitude, to requests for a more "scientific" voice that might be harder to understand but would "feel like it knows things". Ten children asked for a less robot-like voice, while two preferred a clearer distinction between human and machine.

Overall, the children evaluated the system as engaging and enjoyable, but also as inconsistent in its performance. While they appreciated the opportunity to ask questions and receive spoken responses, they were quick to identify breakdowns in understanding, overly complex explanations, and insufficient visual support. As one child summarized, the platform was "sometimes good, sometimes not", capturing both the promise and current limitations of this emerging technology. Notably, one teacher reported that a student expressed a desire to "work with this when I grow up", suggesting that AI-supported learning experiences may have the potential to inspire long-term educational and career aspirations.

## 5. Discussion and limitations

We obtained insights on children's expectations of, interactions with, and evaluations of an LLM-enabled visualization deployed in a classroom to foster conversations around Earth- and space-related datasets. We now contextualize our most salient findings, infer important design guidelines, and note the limitations of our work.

### 5.1. Findings

Prior to interaction, children's expectations about the system were shaped by three interrelated factors: technological novelty, desire for local and personally meaningful content, and school-related norms for educational use. The children expected the platform to be impressive and advanced. Expressions of surprise and excitement suggest that the prospect of speaking to an AI-enabled globe carried a strong "wow-factor," where the perceived intelligence and responsiveness of the system were expected to produce an extraordinary experience. Despite the novelty of the AI component, children largely framed their expectations in educational terms. They anticipated asking "appropriate" school-related questions and expressed frustration when the platform was restricted to a single domain. This suggests that children perceived the platform not merely as a visualization interface, but as a general conversational knowledge agent, consistent with prior research showing that children expect voice-based AI systems to provide broad, flexible informational support [GHD22]. Additionally, children expected the platform to connect to their everyday lives and environments, rather than remaining focused on abstract or distant phenomena such as outer space. Requests to explore familiar places, languages, food, and culturally recognizable content indicate an expectation that the technology should support situated and place-based learning.

During the facilitated sessions, children's engagement with the platform was strongly shaped by adult scaffolding, personal interests, and the technical affordances of the LLM interaction. Children's questions were clearly constrained by the introduction and facilitation. The educator's demonstration questions and explicit framing of the platform's capabilities guided the types of questions children asked, resulting in a predominance of fact-oriented and domain-aligned queries. This aligns with prior findings showing that facilitated use of digital globes tends to structure exploration and focus attention on predefined content rather than open-ended discovery [BEB*15]. The children's prior knowledge also influenced how they engaged with the platform. Although space was foregrounded in the introduction, many children asked questions about Earth-related topics such as population or places. This mirrors findings from both classroom and museum studies showing that children often anchor interaction with complex visualization systems in concepts they already recognize, particularly when learning occurs in institutional settings rather than free-choice environments [SPCH25].

The interaction remained largely non-dialogic. While the conversational interface invited to dialogue, its inability to manage turn-taking, recognize hesitation, or adapt to children's communicative cues resulted in interaction patterns characterized by single-turn questions and answers. Similar breakdowns in conversational flow have been documented in studies of children interacting with voice-based systems, where anthropomorphic cues create expectations of responsiveness that current systems cannot fully meet [AR23]. As a result, children often waited passively for responses, struggled to ask follow-up questions, or experienced misinterpretations when microphone handling or timing failed.

Post sessions, children's evaluations of the LLM-enabled visualization platform combined positive responses with critical assessments of the AI's communicative and explanatory limitations. Overall, the system was experienced as fun and interesting, but also inconsistent and at times difficult to understand. The children's reflections that it was "fun" and that it "showed interesting things" suggest that the platform successfully stimulated curiosity and exploratory engagement. At the same time, the children often evaluated the system in terms of how well it understood them and how clearly it explained things. They noted that the AI did not always grasp their questions, that the responses were sometimes overly long or used difficult words, and that the verbal explanation did not always align with the visual output. Such mismatches reduced perceived competence of the LLM and led children to describe the system as unreliable, or even that it felt like it was "born yesterday."

Voice characteristics of the platform also influenced children's evaluations. Many commented that it "sounded like an AI" due to the lack of emotional variation, breathing, and prosodic nuance, aligning with prior findings that children are sensitive to vocal cues when judging the humanness, competence, and trustworthiness of conversational agents, and that the voice and linguistic style strongly shape how such systems are interpreted, see [DDO*24, GHD22, JSM25]. Importantly, children did not uniformly want the platform to sound like themselves. Several explicitly rejected the idea of a childlike voice, instead expressing preferences ranging from a clearly machine-like voice to a more "scientific" or authoritative tone. This supports earlier findings that, while anthropomorphic cues can increase engagement, they may also cre-ate unrealistic expectations about understanding and responsiveness, particularly in educational contexts [STJ*21].

Taken together, the gained insights point to several implications for the future development of LLM-enabled visualization platforms in school contexts. First, meaningful dialogue requires systems that can better read and respond to children's interactional cues, including turn-taking, hesitation, and follow-up intent. Prior research suggests that current conversational agents lack this interaction awareness, and that museum educators themselves are skeptical that AI systems will soon be able to fully replace human pedagogical judgment in this regard [STJ*21]. Second, children's interaction patterns highlight the importance of allowing the AI to respond flexibly to questions that connect to children's everyday lives and local surroundings, rather than limiting interaction to predefined thematic domains. Studies of AI-supported guides in educational and museum settings similarly emphasize the value of contextualized, situated responses that relate information to visitors' lived experiences [GRS21, STJ*21]. Finally, these findings also underscore the importance of how LLM-enabled visualization platforms are introduced and framed in classrooms. Educators play a central role in shaping children's expectations and guiding which questions are perceived as appropriate, thereby influencing how the technology is used and understood within formal learning environments.

### 5.2. Limitations

Our study aimed to provide early stage information on children's expectations of, interactions with, and evaluations of an LLM-enabled visualizations in the classroom for educational purposes. While our findings add to the nascent body of work on this topic, they suffer from some limitations that are important to discuss.

First, while the spherical display used in this study is part of the *TellUs* project, which focuses on communicating earth-related data to children in their classrooms context [BSS*22, STJ*21], this form factor may have influenced the interactions we have observed. Inherently, spherical displays are likely to influence the spatial organization around the device and perhaps even the kind of interactions that are observed. However, since our study focused on verbal and natural-language interaction with the visualization, we believe that the observations we report are likely to transfer to other displays and that our findings are, thus, likely to be generalizable.

Second, we believe that the novelty effect of both the spherical display and the LLM-based interaction are very likely to influence engagement and perceptions through a double novelty effect (see e.g., [BIAI17]). While this is an important contextualization of our findings, we argue that LLM-based interaction with visualization will still likely remain engaging even after the novelty effect wears off. Indeed, voice-based interaction through smartphones or home speakers has now been existing for years and the generation on which we tested our prototype has already been exposed to such agents. Third, any perception of the LLM may be sensitive to how the LLM itself is implemented: the voice chosen, the vocabulary it uses [JIB*25]. A rigorous study on the impact of different implementation parameters would be needed to properly understand their impact on the perception of an LLM-augmented visualization.

Finally, children used a shared microphone instead of recording everything everyone said and asking the LLM to answer all it perceived. This created a form of turn-taking between the children with the microphone acting as a tangible token for it. The HCI literature highlights that such tangible tokens can create specific embodied dynamics especially in group settings [MFH*09, GMR20]. While this certainly helps to contextualize our findings and perhaps limit their generalizability, we also wish to highlight that classroom settings with young audiences requires proper organization and turn-taking to constrain behaviors, and that the microphone also helped us avoid facing technical challenges for the LLM to properly understand what the children were saying.

With these limitations considered, we still argue that our findings are important and generalizable enough to provide interesting design guidelines for future research in this domain and to highlight several important future avenues for research.

### 6. Conclusion

In this paper, we presented a classroom study of an LLM-enabled spherical visualization platform designed to support children's engagement with earth and space-related datasets. We particularly focused on children's expectations prior to interaction, their engagement during facilitated sessions, and their reflections afterward. Our findings suggest that, while the combination of conversational AI and shared immersive visualization can stimulate curiosity and engagement, it also generates strong expectations regarding dialogic capability, explanation clarity, and alignment between verbal and visual responses. The results highlight the central role of educator's facilitation, framing, and system transparency in supporting children's interpretation and evaluation of LLM-enabled visualizations in classrooms. This work thus contributes early observations to the emerging space of LLM-augmented educational visualization and identifies potential design considerations for future systems. Our intention is that these exploratory findings will help inform further technical development and future educational research on how LLMs can be meaningfully integrated into visualization tools for learning in classrooms, as well as public contexts at large.

### 7. Acknowledgments

This work was funded by three research grants: Marcus and Amalia Wallenberg Foundation (2023.0128) and (2023.0130), and Knut and Alice Wallenberg Foundation (2019.0024). We wish to thank the school and children who participated in the study.

### 8. CRediT authorship contribution statement

**E.F**: Conceptualization; Data Curation; Resources; Formal analysis; Writing – original draft; Writing – review & editing. **I.S**: Conceptualization; Data Curation; Formal analysis; Resources; Writing – original draft; Writing – review & editing. **M.B**: Software; Data Curation; Resources; Writing – original draft; Writing – review & editing. **A.G**: Conceptualization; Writing – review & editing. **K.S**: Conceptualization; Funding acquisition; Writing – review & editing. **A.L**: Conceptualization; Writing – review & editing. **M.F.J.**: Software; Writing – review & editing. **E.S**: Software; Writing – review & editing. **Y.S**: Resources; Data Curation. **A.Y**: Conceptualization; Funding acquisition; Writing – review & editing. **M.R**: Conceptualization; Writing – review & editing. **L.B**: Conceptualization; Funding acquisition; Writing – original draft; Writing – review & editing.

10 of 10 *Emelie Fälton et al. / Children's Expectations, Engagement, and Evaluation of an LLM-enabled Spherical Visualization Platform in the Classroom*[OSH19] OTTERBORN A., SCHÖNBORN K., HULTÉN M.: Surveying preschool teachers' use of digital tablets: general and technology education related findings. *International journal of technology and design education 29*, 4 (2019), 717–737. doi:https://doi.org/10.1007/s10798-018-9469-9. 2, 3

[PSN24] PAPANTONIS STAJCIC M., NILSSON P.: Teachers' considerations for a digitalised learning context of preschool science. *Research in Science Education 54*, 3 (2024), 499–521. doi:https://doi.org/10.1007/s11165-023-10150-5. 2

[PWD*24] PATRON E., WERNHOLM M., DANIELSSON K., PALMÉR H., EBBELIND A.: An exploration of how multimodally designed teaching and the creation of digital animations can contribute to six-year-olds' meaning making in chemistry. *Education Sciences 14*, 1 (2024), 79. doi:https://doi.org/10.3390/educsci14010079. 2

[RBZ*25] REINDERS S., BUTLER M., ZUKERMAN I., LEE B., QU L., MARRIOTT K.: When refreshable tactile displays meet conversational agents: Investigating accessible data presentation and analysis with touch and speech, 2025. arXiv:2408.04806, doi:https://doi.org/10.1109/TVCG.2024.3456358. 3

[SB24] SCHÖNBORN K., BESANÇON L.: What can educational science offer visualization? a reflective essay. In *2024 IEEE VIS Workshop on Visualization Education, Literacy, and Activities (EduVIS)* (2024), pp. 30–37. doi:10.1109/EduVIS63909.2024.00009. 2

[SPCH25] SHABY N., PELEG R., COOMBS I., HEMMING J.: "it was just amazing!" unstructured interactions following a planetarium and science shows. *Journal of Museum Education* (2025), 1–15. doi:https://doi.org/10.1080/10598650.2025.2461846. 7

[SSE*22] SATRIADI K. A., SMILEY J., ENS B., CORDEIL M., CZAUDERNA T., LEE B., YANG Y., DWYER T., JENNY B.: Tangible globes for data visualisation in augmented reality. In *Proceedings of the 2022 CHI Conference on Human Factors in Computing Systems* (New York, NY, USA, 2022), CHI '22, Association for Computing Machinery. doi:10.1145/3491102.3517715. 2

[STJ*21] SONI N., TIERNEY A., JURCZYK K., GLEAVES S., SCHREIBER E., STOFER K. A., ANTHONY L.: Collaboration around multi-touch spherical displays: A field study at a science museum. *Proc. ACM Hum.-Comput. Interact. 5*, CSCW2 (Oct. 2021). doi:10.1145/3476067. 2, 3, 8

[SUAO*14] SCHOLLAERT UZ S., ACKERMAN W., O'LEARY J., CULBERTSON B., ROWLEY P., ARKIN P. A.: The effectiveness of science on a sphere stories to improve climate literacy among the general public. *Journal of Geoscience Education 62*, 3 (2014), 485–494. doi:https://doi.org/10.5408/13-075.1. 1, 2, 3

[VWB*14] VEGA K., WERNERT E., BEARD P., GNIADY C., REAGAN D., BOYLES M. J., ELLER C.: Visualization on spherical displays: Challenges and opportunities. In *Proceedings of the IEEE VIS* (2014), pp. 108–116. 2

[WSB15] WILLIAMSON J. R., SUNDÉN D., BRADLEY J.: Globalfestival: evaluating real world interaction on a spherical display. In *Proceedings of the 2015 ACM International Joint Conference on Pervasive and Ubiquitous Computing* (New York, NY, USA, 2015), UbiComp '15, Association for Computing Machinery, p. 1251–1261. doi:10.1145/2750858.2807518. 2
© 2026 The Author(s).
Proceedings published by Eurographics - The European Association for Computer Graphics.